# An Explanation of Dayton Miller's Anomalous "Ether Drift" Result

Thomas J. Roberts

*Illinois Institute of Technology, Chicago, IL.*
*and*
*Muons, Inc., Batavia IL.*

**Abstract**
In 1933 Dayton Miller published in this journal the results of his voluminous observations using his ether drift interferometer, and proclaimed that he had determined the "absolute motion of the earth". This result is in direct conflict with the prediction of Special Relativity, and also with numerous related experiments that found no such signal or "absolute motion". This paper presents a complete explanation for his anomalous result by: a) showing that his results are not statistically significant, b) describing in detail how flaws in his analysis procedure produced a false signal with precisely the properties he expected, and c) presenting a quantitative model of his systematic drift that shows there is no real signal in his data. In short, this is every experimenter's nightmare: he was unknowingly looking at statistically insignificant patterns in his systematic drift that mimicked the appearance of a real signal. A value of zero with an upper limit on "absolute motion" of 6 km/sec is derived from his raw data, fully consistent with similar experimental results and the prediction of Special Relativity. The key point of this paper is the need for a comprehensive and quantitative error analysis. The concepts and techniques used in this analysis were not available in Miller's day, but are now standard. These problems also apply to the famous measurements of Michelson and Morley, and to most if not all similar experiments; appendices are provided discussing several such experiments.

## CONTENTS

## I. Introduction

Dayton C. Miller's 1933 article in this journal reviewed the results of his voluminous measurements obtained from his "ether drift" interferometer, and proclaimed to the world that he had determined "the absolute motion of the earth". This claim has been embraced by some, rejected by many, and remains controversial today[1]. Unfortunately, acceptance or rejection of Miller's result has been based primarily on whether or not it conforms to a person's prejudices, and not on any solid, objective criteria. Shankland *et al.* (1955) attempted to address this situation by re-examining Miller's data. Unfortunately they did not fully resolve the issue, because they merely showed a loose correlation between signal and temperature drift, but did not give any argument or discussion of how that could generate such a remarkable result. Posthumously, Miller is now experiencing every experimenter's nightmare: the standard analysis algorithm used by himself and all his contemporaries had serious flaws not explicable for many decades, and they caused the noise in the apparatus to perfectly mimic the signature of a real signal. This paper provides a detailed and complete explanation of his anomalous result, giving solid and credible reasons to reject Miller's claim: a) a basic error analysis of his results shows that they are not statistically significant, b) his analysis algorithm has serious flaws that force the noise in his data to precisely mimic the properties he expected in a real signal, and c) a quantitative model of his systematic drift accounts for 100% of his usable data, leaving no real signal. This re-analysis of his data derives a value of zero with an upper limit of 6 km/s (90% confidence level) on the "absolute motion of the earth", which is fully consistent with related experiments[2] and the prediction of Special Relativity.

---

[1] See, for example, the papers by Allais (1998, 1999a, 1999b, 2000), Cahill (2002, 2003a, 2003b, 2004, 2005), Cahill and Kitto (2002), Consoli (2003, 2005), Consoli and Costanzo (2003a, 2003b, 2004), Deen (2003), DeMeo (2001), Munera (1997, 1998), Sato (2006), Selleri (2000), and Vigier (1997). These are all dissident authors, and they all build upon the assumption that Miller's results are valid; none of these authors include a comprehensive error analysis, including the few who claim to do so (every one ignores the huge systematic drift, and none performs the simple error analysis given here in section II). The mainstream of physics assumes his results are invalid. Neither group has had solid, objective criteria to support their position, until now.

In addition, appendices are provided which apply this type of analysis to the measurements of Michelson and Morley (1887), and also to the measurements of Illingworth (1927). These quite similar experiments are also plagued by the same problems, but because their systematic errors were much smaller than Miller's, these authors contented themselves with merely claiming an upper bound on any signal – such upper bounds remain valid, and this analysis explains why those upper bounds are larger than their observational resolution would suggest.

Miller's interferometer is a direct descendant of Michelson's original version, and Michelson and Morley's improved model – indeed he reused their mercury trough in which the interferometer floats while rotating. But instead of using a sandstone block for optical and thermal rigidity, iron girders were used to increase the optical path. Miller (1933) has several diagrams and pictures of the apparatus. Miller streamlined the data acquisition process by having the observer visually interpolate the fringe position, rather than using a vernier dial and readout. And he took data in runs of 20 turns (rather than 6 turns), a seemingly minor change that turns out to be important because it yields sufficient data to permit the quantitative modeling of the systematic drift in section IV. Miller clearly appreciated the desirability of having a large data set.

This paper first discusses Miller's data reduction algorithm, including an error analysis of that algorithm, showing that the errorbars are enormous and his stated results are not statistically significant. Then a discussion of his data analysis in the frequency domain is presented using digital signal processing (DSP) techniques, showing that his algorithm forces the noise that is present to mimic the exact type of signal he expected. Next a re-analysis of his data is given, in which a direct quantitative model of his systematic drift is shown to account for 100% of the usable data, leaving no real signal. And finally some conclusions are presented. Appendix I discusses the famous measurements by Michelson and Morley, and Appendix II discusses the measurements of; these problems are related to the structure of the analysis algorithm, and similar conclusions could reasonably be expected for all experiments of this type analyzed using an algorithm similar to Miller's.

Miller himself could have had no knowledge of these DSP techniques developed after his death, and without a digital computer the computations presented here would have greatly exceeded the already heroic efforts he made to acquire the data. The key point of this paper is the need for a comprehensive and quantitative error analysis and the importance of errorbars; this is well known today, but was not common in Miller's day[3].

## II. Miller's data reduction algorithm

Miller took his data and analyzed them as follows: The interferometer was set rotating slowly and was adjusted so the fringes were visible with the center fringe near the center of the field of view. There is a small pointer affixed to one of the mirrors, and the central fringe position is visually measured relative to it, in tenths of a fringe width. The observer walked slowly around a circle watching the image in the telescope. There are 16 markers spaced equally around the circumference, and the observer called out the position of the central fringe at each marker. An assistant recorded the values, with each turn starting and ending at Marker 1, so a single Marker 1 reading was recorded twice for successive turns (complete rotations). Whenever the central fringe drifted more than about 2 fringes from the pointer, small weights were applied to one of its arms to flex the interferometer and bring the center fringe back near the pointer. These adjustments were made during an unrecorded turn and always appear in the logbook between the Marker 1 value at the end of one turn and the Marker 1 value at the beginning of the next turn. Typically the 20 recorded turns of a run took about 15 minutes, and the interferometer had enough inertia to rotate on its own with little change in rotation rate during this time. Marker number 1 was always oriented to the North, and is the start and end of each turn, so the markers designate fixed orientations of the interferometer relative to the lab.

---

[2] A summary of similar experiments is in Table I of (Shankland, 1955); a more recent and wider discussion of tests of Special relativity is in (Will, 2005), and a larger compilation of experimental tests is in (Roberts, 2006).

[3] Indeed, in the entire volume in which his paper appeared not a single plot displays errorbars.

The fringe positions at each marker are recorded in the logbook as shown in Fig. 1, taken directly from (Miller, 1933). After the header giving basic information there are twenty rows of 17 columns of fringe positions (recorded in tenths of a fringe); column 17 is the marker 1 value at the end of the turn and is equal to the column 1 value of the next turn except when adjustments were made between recorded turns. Below the 20 rows of raw data from 20 turns are ten rows of manual computations that a) sum the first 20 rows, b) subtract the linear drift between columns 1 and 17, c) divide by 20 to yield an average, d) subtract the mean of the 16 values, and e) average the first and second ½ turns. Below the computations are plots of the reduced full-turn and ½-turn results. Figure 1 uses units of 0.1 fringe throughout, and the amplitude of the final ½-turn plot is about 0.06 fringes. This algorithm is essentially identical to that used by Michelson (1887) and by all contemporary interferometer measurements (the differences are in the number of turns taken during each data run, and some authors force the endpoints to zero rather than forcing the mean to zero).

Miller expected the real signal to be a sinusoid with a period of ½ turn, with its amplitude related to the speed of absolute motion and its phase related to the direction. The ease of seeing a "signal" in these data is illustrated in Fig. 1, which shows a plot of his "signal"[4] at the bottom. That plot quite clearly shows a reasonably sinusoidal variation with a period of ½ turn, and it certainly looks like Miller expected a real signal to look. But that is a plot without errorbars, and one cannot determine if that variation is statistically significant or not, so we must consider the errors in this measurement.

The first thing to do is look at the data of Fig. 1, plotted in Fig. 2, which shows the raw data with adjustments restored. There is huge variation about 100 times larger than the amplitude of the plot at the bottom of Fig. 1, and extracting any signal from such a large background is a challenge. Clearly the interferometer systematically drifted about 6 fringes during this run, as the large-scale changes cannot possibly be any real signal with a period of ½ turn. From the way data were recorded and from Miller's own estimates, the statistical errors in these data are on the order of 0.1 fringe. So both the size and the shape of the variations in Fig. 2 imply that this is a large systematic error: for instance, all measurements in the last turn are systematically almost 6 fringes smaller than those in the first turn. The adjustments after turns 5, 9, and 19 make a crude subtraction of this large drift, but even so the averaging of the turns cannot remove the effects of this drift because it systematically introduces correlations in the values.

Miller accounted for this drift by subtracting the linear drift between Marker 1 and Marker 17 from his averaged data. Today this would be called subtracting a model of the systematic drift, and this model simply assumes it is linear. While Miller subtracted the average linear drift after averaging the turns, this is equivalent to subtracting the linear drift of each turn individually, and it is instructive to consider it that way. The lines in Fig. 3 display this linear model for the systematic drift. The corners of the lines are at successive Marker 1 values, and the individual points are the values at Marker 9, 180 degrees away from Marker 1. Because of the 180 degree symmetry of the interferometer, any real signal can depend only on orientation modulo 180 degrees, and whatever real signal is present in these data must have the same value for every reading at Markers 1 and 9. So the effect of a real signal is to move the entire plot vertically by some constant amount; all of the variations in Fig. 3 are purely instrumentation effects, i.e. a systematic drift.

But the full value of Fig. 3 is in comparing the points to the lines to test the assumption of linearity. Since any real signal corresponds to a constant vertical offset in the entire plot, if the systematic drift were truly linear all of the points would lie on the lines. Clearly that is not so, and some points miss the lines by as much as 0.5 fringe. So the assumption that the systematic drift is linear is not valid to better than about 0.5 fringe – that is almost a factor of 10 larger than the variations in the plot at the bottom of Fig. 1. Of course this is merely a quantitative statement of what the eye can clearly see – the systematic drift in Fig. 2 is clearly not at all linear during many if not most of the turns.

---

[4] As discussed in (Miller 1933) a harmonic analyzer was used to extract the amplitude of the ½-turn Fourier component from plots like this; in this paper a discrete Fourier transform (DFT) is used.

While Fig. 3 shows the inadequacy of assuming a linear drift, it is still useful to obtain quantitative errorbars for these data analyzed in this manner, because this is what Miller used to obtain his results. A general rule for error analysis is that when averaging raw data to obtain a value one can histogram the data to obtain the corresponding errorbar. Each point in the plot at the bottom of Fig. 1 consists of the average of 40 readings (two marker values 180 degrees apart for each of 20 turns of the interferometer). The histogram of the data for Markers 1 and 9 is shown in Fig. 4; it has a sigma of 0.7 fringe. In addition to the histogram, the figure has each contribution labeled by the turn of each data point (A-T correspond to turns 1-20). The other seven histograms corresponding to the other orientations are similar, and all have sigmas of 0.7 to 0.8 fringes. If these 40 measurements were statistically independent one would divide the sigma of the histogram by $\sqrt{40}$ to obtain the errorbar for the mean. But if the error were statistical, then the contributions would fill the histogram in random order; this is clearly not the case, and symbols A-E and O-S march systematically from right to left – such patterns clearly indicate this histogram is of a systematic drift that forces the value to "scan" across the histogram. This is also clear from a look at Fig. 2, and for such systematic effects the sigma of the histogram corresponds to the errorbar of the average. Such patterns and the obvious drift suggest that a model of the systematic drift could be used to greatly reduce the errorbar, and Section IV below does just that. Figure 5 shows the ½-turn plot at the bottom of Fig. 1 with these errorbars. Clearly the variations are not significant.

Since Miller's day our attitude toward experiments like this has changed, and we now use them to test theories, rather than to "determine the absolute motion of the earth". And we do this quantitatively using $\chi^2$ fits (or similar). So the modern approach to interpreting these data in Miller's theoretical context would be to start with Miller's model of absolute motion as applied to his instrument, and test the class of theories "The earth is moving with absolute speed X in direction Y" where X and Y are determined by fitting to the data. The speed X is related to the amplitude of the signal, and the direction Y is related to its phase. Miller's conversion from signal amplitude to absolute speed is given in Fig. 20 of (Miller, 1933), in which 0.7 fringe corresponds to 24 km/sec. Looking at Fig. 5, it is clear that this run will have a good $\chi^2$ for any sine wave with amplitude corresponding to speed X less than about 30 km/sec and phase corresponding to any direction Y whatsoever. So the errorbars on X and Y are huge. This is just one run out of hundreds, and some have smaller errorbars, some have larger errorbars. But all runs in the data sample have the property that the errorbars exceed the variation in the final ½ turn plot, as in Fig. 5. That means that this analysis cannot really determine the direction of absolute motion at all, and cannot say very much about the speed other than that it is less than about 30 km/sec. (Miller, 1933) displays several plots of absolute speed and direction, but they are all without errorbars. Had he computed and plotted errorbars as above, they would be so large that in no case would they fit on the plot, and often would not even fit on the page. His "determination of the absolute motion of the earth" is not statistically significant. Because of the flaws in this data reduction algorithm (discussed next), there's no point in actually performing a detailed statistical analysis of results from this analysis method.

## III. Miller's analysis in the frequency domain

Considerable insight into the effects of the data analysis procedures is obtained by considering a Fourier transform of the data and procedures. The data were manually sampled at a rate of 16 samples per turn with a resolution of 0.1 fringe (in modern terms, the human observer was performing the function of a sampling analog to digital converter). Figures 1 and 2 contain 20 complete turns of 16 samples each, with a total of 321 samples (one extra reading at Marker 1 fills out the 20[th] row of Fig. 1).

Figure 6 displays the frequency spectrum[5] of the data of Fig. 2, obtained via a 320-point discrete Fourier transform (DFT). Low frequency components clearly dominate the spectrum, and it is reasonably close to the spectrum of 1/f noise. There is a small bump in frequency bin 40, which corresponds to a period of ½ turn, and any real signal would be in that bin. We'll come back to it.

---

[5] As usual, only the norm of the complex value of each frequency bin is shown, and only the lower half of the bins is shown because the upper half of the bins are merely copies of these (aliased by the sampling).

The first analysis step is to sum the 20 turns, marker by marker. For descriptive purposes it is convenient to combine this with the later division by 20 so it is an average of the 20 turns. In the frequency domain this averaging is a comb filter[6], with a frequency response shown in Fig. 7. Except for the dc component in bin 0, the low-frequency components (periods longer than 1 turn) are eliminated by this filter, as are higher components that are not harmonics of 1 turn. After applying this filter the data have only 8 nonzero frequency bins, and can be reduced to 16 samples without loss of information (of course this reduction is the natural result of averaging); in the frequency domain this corresponds to the eight non-zero frequency bins of Fig. 7 becoming the 8 frequency bins of the 16-point DFT spectrum. This 16-point DFT spectrum of the comb-filtered data is shown in Fig. 8. In reducing the number of samples, the ½-turn signal bin has moved to bin 2, and of course remains 0.11 fringes.

The next analysis step is to subtract the linear systematic model from the 16 comb-filtered data points. For the Fig. 1 data this is a ramp from 0 to 0.305 fringes, reducing the bin 2 amplitude to 0.06. Next the mean of the data is subtracted, removing the dc component in frequency bin 0. And finally the two halves of the 16 point 1-turn signal are averaged to an 8-point ½-turn signal. That is another comb filter that retains only the even-numbered frequency bins, giving the final spectrum shown in Fig. 9; the ½-turn signal bin is now number 1. Note that only 4 frequency bins remain, and one of them (bin 0) has been explicitly zeroed.

A conspicuous feature of these spectra is that they all have decreasing amplitude with increasing frequency. And in the final plot the frequency bin in which the real signal would appear is bin 1, the lowest nonzero frequency bin. From the nature of Fourier analysis, it should be clear that after this analysis any initial data with a falling spectrum (e.g. 1/f noise) will have a spectrum similar to Fig. 9, and a time domain appearance similar to the plot at the bottom of Fig. 1. This is a simple consequence of the fact that the ½-turn Fourier component is the lowest frequency retained by the algorithm, and it will dominate because of the falling spectrum. When a single frequency bin dominates the Fourier spectrum, the signal itself looks approximately like a sinusoid with that period. Using this data reduction algorithm, any noise with a falling spectrum will end up looking like an approximately sinusoidal "signal" with a period of ½ turn – precisely what Miller was looking for. Figure 5 shows that this data run does indeed have a lot of noise with a falling spectrum, so the shape of the plot at the bottom of Fig. 1 does not necessarily imply that there is a real signal present – one needs a more sophisticated analysis to determine that (see section IV).

There are three basic flaws in this data reduction algorithm: a) averaging the data, b) assuming the systematic drift is linear, and c) absence of a quantitative error analysis. These were not considered flaws in Miller's day, and all experiments of this type were analyzed with minor variations of this algorithm. Together with his rather large systematic drift, however, they permitted Miller to be fooled into thinking there is a real signal present. In the next section a new analysis is presented which avoids all of these flaws.

In Miller's day, the ideas used in the above discussion were not well known or fully developed, and it's not surprising that he was fooled. Today we know that the averaging of data during analysis should be avoided whenever possible, and one would of course use the full 320-point DFT of Fig. 5. That spectrum clearly shows that the low-frequency bins (corresponding to a slow systematic drift) completely dominate any real signal. Unfortunately, this approach is not able to determine whether the content of frequency bin 40 (period ½ turn) is a real signal or is due to the systematic drift. Doing that requires an accurate model of the systematic drift, discussed next.

# IV. A re-analysis of Miller's data

The basic challenge of analyzing these data is dealing with the very large systematic drift. Fortunately Miller took enough data so that it can be well modeled for most of his data runs. The basic technique is to analyze each run separately, modeling the run's data as a sum of a periodic signal plus a systematic drift. Any real signal, of course, has a dependence on orientation (marker #) that is independent of time (turn #) during the run. So the data are modeled as the sum of a signal and a systematic drift:

---

[6] See any textbook on digital signal processing, such as (Rabiner, 1975).

```
    data = signal(orientation) + systematic(time)
```

The key point is that signal(orientation) is independent of time, and for each orientation (marker) it has the same value for every turn of the interferometer within a given data run[7]. Therefore if the data from the first turn is subtracted marker-by-marker from the data of every turn, the result is completely independent of any orientation dependence, and contains only systematic(time). It is convenient to take advantage of the 180 degree symmetry of the apparatus, and combine the data for markers 180 degrees apart. This gives 8 orientations, and 8 independent measurements of differences in systematic(time), shown in Fig. 10 – each measurement is a sequence of values at a given orientation. Because the values from the first ½ turn were subtracted from the values of each orientation, any real signal has been removed from the plot. Each of the eight measurements of the systematic drift starts with a 0 for the first ½ turn, and we don't know *a priori* how to combine them into a single function corresponding to systematic(time).

Because those eight sequences are interleaved in time by the rotation of the instrument, if we assume that the variations in systematic(time) are as small as possible we can fit the eight sequences to a single function of time. Eight parameters are introduced, the initial values of each orientation's measurement of systematic(time), and each parameter is added to the values of the 40 points for its orientation. This corresponds to permitting each sequence of Fig. 10 to move vertically, independent of the others, and the fit will determine the best positions for the eight measurements such that they combine as smoothly as possible into a single continuous function systematic(time). The $\chi^2$ is formed from the differences between adjacent points in time, summed over all 320 differences (all turns, all orientations). To avoid the usual normalization ambiguity in such a fit[8], the first parameter is fixed by requiring the model to match the first data point, leaving seven free parameters. As the data are quantized at 0.1 fringe, so are the parameters, and instead of the usual minimization programs an enumeration of all reasonable sets of parameters was used with an algorithm that finds the minimum $\chi^2$. The result of the fit is a complete quantitative model of systematic(time) for the run. This fit has 313 degrees of freedom, and the histogram of $\chi^2$ for all runs has a mean of 300, indicating that the estimate of the individual measurement resolution (0.1 fringe) is reasonable. Fitting each run took about 3 minutes of computer time to enumerate several million combinations of the 7 parameters to find both the best fit and the errorbar on the ½ turn Fourier amplitude.

This analysis has been applied to a sample consisting of 67 of Miller's original data runs, including runs from his 1925, 1927, and 1929 Cleveland data and his 1925-26 Mount Wilson data. They were selected to span all available epochs of his data (months of data taking) and all sidereal times, without regard to content.

These data display frequent instabilities, including occasional drifts of more than 2 fringes per turn and occasional jumps as large as 1.5 fringes between successive markers (!). One run drifted by 18 fringes in 17 turns; but then, three runs drifted by less than 1 fringe during 20 turns. Examination of the data suggests classifying regions of instability as any drift with a rate of more than ½ fringe during ½ turn. That is five times the largest signal Miller claimed, and almost ten times the amplitude of the plot at the bottom of Fig. 1. These regions of instability do not display any consistent orientation dependence in any run. The model of the systematic drift cannot be expected to fit runs with major instabilities, because its assumption that the systematic drift is as small as possible is violated for such rapid drifts. The presence of instabilities has a modest effect on the $\chi^2$ of the fits to the systematic drift, but the presence of a large number of instabilities in a run has a significant effect on whether or not the systematic model matches the data. The best characterization found is the total number of turns in a run without instabilities: the plot of the raw data for each of the 67 runs in the sample was manually examined, separating it into regions of instability (during which the interferometer drifted at least ½ fringe during ½ turn), and regions of stability (1 turn or more with no such rapid drift). The durations of the stable regions were then summed, without requiring them to be consecutive.

Figure 11 shows the results of this analysis, displaying the ½-turn Fourier amplitude of data minus systematic for all 67 runs in the sample; this is the true signal for this analysis. Runs plotted with closed circles have at least 6 stable turns, while those plotted with open circles have 5 or fewer stable turns (i.e. ¾ or more of the turns are unstable by the above criterion). Each fit includes the entire run; the instabilities are used only to select an open or closed circle for the plot. All of the closed circles have zero amplitude, because the systematic model reproduces the data exactly for all runs with moderate or good stability. The lack of variance in these runs' amplitudes is explained by the quantization of both data and parameters at 0.1 fringe; varying any one parameter by its quantum increases $\chi^2$ significantly. The runs which have nonzero amplitude all have major instabilities of the instrument throughout the run, and the plot of data-systematic is zero except for one point of 0.1 fringe (the quantum) or two points of 0.1 fringe with the same sign – these cannot reasonably be interpreted as a "sinusoid signal", and look much more like a fit to a systematic drift that is not as small as possible.

---

[7] The variation due to the rotation of the earth is negligible, because during a run it is less than 6% of the angle between markers, and corresponds to a measurement timing error of less than 0.2 seconds, smaller than the reaction time of the human observer.
[8] The $\chi^2$ is made up of differences, so any constant can be added to all 8 parameters without changing $\chi^2$.

It is reasonable to omit all runs with instabilities during ¾ of the run, as the systematic model cannot be expected to model instabilities very well – its assumption that the systematic is as small as possible is violated by such rapid drifts. Indeed it is surprising that such a large fraction of unstable turns is accommodated. There are 14 runs that fail this cut (21%), and of those only 9 runs (13%) have nonzero Fourier amplitude with period ½ turn. These runs have no obvious dependence on sidereal time, and come from several data epochs. The conclusion is that during these runs the interferometer was just too unstable for the systematic drift to be modeled well. Without an accurate model of the systematic drift any analysis is useless, especially for runs during which the systematic drift was exceptionally large or ill behaved, so these 14 runs have been omitted from further analysis.

The 53 remaining runs are plotted in Fig. 11 with closed circles, and all have the property that the systematic model fit to the systematic measurements is exactly equal to the original data. That is, signal(orientation) is identically zero for each of these runs.

The requirement of at least 6 stable turns per run is rather arbitrary, based on examination of the data, and it is a bit surprising that just 6 stable turns out of 20 is enough for the systematic drift to be modeled well. The conclusions of this analysis are not affected if the requirement is increased to 10 stable turns per run – that merely reduces the number of remaining runs to 42. Requiring all 20 turns to be stable still leaves 13 runs.

The errorbars in Fig. 11 are due to the uncertainty in fitting the systematic model to the measurements of the systematic drift, and correspond to the effect on the ½-turn Fourier amplitude for a unit increase in $\chi^2$, as usual. As the systematic model reproduces the data exactly for these runs, the uncertainty in the systematic fit gives the complete errorbar for this analysis; it inherently includes both systematic and statistical errors. These errorbars are very much smaller than the errorbars of Miller's analysis method discussed in Section II because here the errorbar is due to the uncertainty in a highly constrained fit, while for his method the entire variation of the systematic drift contributes to the errorbar. The results of the two analyses are also quite different, and the errorbar here is ¼ as large as the false signal Miller found at the bottom of Fig. 1 (it is now known to be false because that run passes the instability cut and signal(orientation) is identically zero).

As this analysis concludes there is no signal in Miller's data, it can set an upper limit on any signal with period ½ turn and on the "absolute motion of the earth". From Fig. 11 (omitting runs with open circles as discussed above), a reasonable estimate of the overall errorbar is 0.015 fringe for all sidereal times and all epochs of data. The errorbar is dominated by the systematic drift, and the availability of many runs does not decrease it. That implies an upper limit of 0.025 fringe at the 90% confidence level (1.65 σ). This must then be increased by 1/cos(latitude) to account for the worst-case projection onto the plane of the interferometer. Figure 20 of (Miller, 1933) relates fringe shift to his model of absolute speed, yielding an upper limit on the earth's absolute motion of 6 km/sec (90% confidence level). This value is consistent with similar experimental measurements, and with the null result predicted by Special Relativity.

## V. Conclusion

Dayton Miller was a prisoner of his time. In the 1920s and 30s digital signal processing was unknown, and the serious flaws of the data reduction algorithm used by all such experiments went unnoticed. Also, the use of errorbars and quantitative error analyses were in their infancy. These aspects of the state of scientific knowledge combined to permit him to be fooled into thinking his interferometer measurements did indeed determine the "absolute motion of the earth". Even in 1955, Shankland *et al* did not have knowledge of these aspects of Miller's analysis.

Today, of course, digital signal processing is well known, digital computers are ubiquitous, and quantitative error analyses are presented in essentially all scientific publications. The above discussions of Miller's analysis and data are simply applications of now-standard techniques to his rather ancient data. This paper does not break any new ground, it merely explains a longstanding puzzle: how could someone as competent as Dayton Miller obtain results so inconsistent with other experiments? – As discussed above, he was a victim of every experimenter's nightmare, and was unknowingly looking at statistically insignificant patterns in his systematic drift that mimicked the appearance of a real signal. So it's not surprising that his results were anomalous.

This paper has not only explained how Miller was fooled, it has also presented a re-analysis of his data. This new analysis obtains a value of zero, and puts an upper bound on the "absolute motion of the earth" of 6 km/s (90% confidence level). This is fully consistent with similar measurements, and with the null result predicted by Special Relativity.

We are all prisoners of our time. While this paper gives solid and credible reasons to reject Miller's result, it is unfair to attempt to judge him by the standards of today. Indeed, recognizing that he could not possibly have known about these flaws in his results permits us to admire him all the more for his dedication and perseverance in pursuing these measurements. His instinct that more data is better was correct, and permitted the quantitative modeling of his systematic drift.

## Acknowledgements

Thanks to the Case Western Reserve University Archives for making copies of Miller's data sheets available. And special thanks to Mr. Glen Deen for assistance in transcribing many of those data sheets into computer-readable form.

## Appendix 1. The Michelson and Morley experiment

Michelson and Morley (1887) analyzed their data using a data reduction algorithm quite similar to Miller's, and therefore their result suffers from the same serious flaws discussed above. They did, however, have a smaller systematic error, and they contented themselves with putting an upper bound on the earth's speed relative to the ether of 7.5 km/s.

Unfortunately, their original data have been lost[9], and the only available data are the averages for six data runs in their 1887 paper. So while the modeling of their systematic drift is not possible, it is possible to estimate their errorbars using the same technique as above. Figure 12 displays their reported data, with errorbars computed from a histogram of the data after: a) subtracting the assumed-linear systematic for each run, and b) subtracting the mean for the orientation of each data point. Subtracting the mean for each orientation removes any real signal from the histogram so that all orientations could be combined into a single histogram to improve the poor statistics. Note this errorbar is an *under* estimate, because it comes from the variance of the per-run averages for the markers rather than from the variance of the raw data themselves. It is also an average for the eight orientations. These data are also dominated by the systematic drift, and the sigma of the histogram was used for the errorbars. While it is not possible to draw Fig. 3 for their raw data, one can compare the full-turn assumed-linear systematic for each data run to the marker in the middle: for 3 out of the 6 runs that difference is larger than the errorbars displayed in Fig. 12, so it is clear that their systematic drift is nonlinear by an amount considerably larger than the variations in their data. Just as for Miller's data, it is inadequate to assume that the systematic drift is linear.

While it is not fruitful to attempt a more detailed analysis, it is clear from Fig. 12 that there is no statistically significant signal in their data (remember the errorbars are under estimates). Handschy (1982) comes to a similar conclusion via a different route.

---

[9] Writing in the same building in which their experiment was performed, Miller wrote in a logbook in the 1920s, "The original notes seem not to have been preserved."

# Appendix 2. The measurements by Illingworth

Illingworth's paper (1927) illustrates how a full understanding of analysis techniques is very important when designing an experiment. He took data at only four points around the circle, so for a signal with a period of ½-turn his data are under-sampled, and one cannot extract the Fourier amplitude from his measurements. Had he known this, he would surely have taken data at more points, but of course Shannon's sampling theorem was then 20 years in the future.

The innovation in this interferometer is a 1/20 wavelength step in one mirror that permits the observer to take readings accurate to 0.002 fringe, very much better resolution than any other experiment of this type. The interferometer is stopped at each of four orientations per turn and the observer adjusts the instrument with small weights to center the image; the number of weights was calibrated to give the reading for fringe shift. His use of a constant-temperature room also reduced his systematic drift considerably compared to the other experiments.

Because of the under sampling it is not possible to perform useful Fourier transforms, nor is it possible to model the systematic drift. But it is worthwhile to examine the data, compute errorbars, and determine whether or not there is any significant variation in the data. Figure 13 displays the one run for which data are available from Table II in Illingworth (1927). Illingworth did not plot the data, but did perform averages and subtract an assumed-linear systematic in his analysis, so in Fig. 13 the data points are the per-orientation averages of the ten turns minus the straight line between the average North measurements. The errorbar for each orientation came from the sigma of the histogram of its ten readings divided by $\sqrt{10}$, because it is not clear if the errors are systematic or statistical; as there is likely some systematic component of the error, these errorbars are probably under estimates. The interferometer was adjusted to zero before the start of each turn, so the initial reading at North has a zero errorbar (this also makes determining the overall systematic drift impossible). Comparing the individual turns' linear systematic to the value at the midpoint (as in Fig. 3) shows that four of the ten turns had a nonlinearity in the systematic that exceeds the errorbars displayed in Fig. 13, so the assumption that the systematic is linear is inadequate here, too. In any case, clearly there is no significant variation in these data.

# REFERENCES

Allais, M., 1998, "The Experiments of Dayton C. Miller (1925-1926) and the Theory of Relativity", 21st Century, Spring 1998, p26-32 (1998).
Allais, M., 1999a, C. R. Acad. Sci., Paris, t. 327, Serie IIb, p1405-1410 (1999).
Allais, M., 1999b, C. R. Acad. Sci., Paris, t. 327, Serie IIb, p1411-1419 (1999).
Allais, M., 2000, "L'orgne des regularites constatees dans les observations interferometriques de Dayton C. Miller 1925-1926: variations de temperature ou anisotropie de l'espace", C. R. Acad. Sci., 1, Serie IV, p1205-1210 (2000).

Cahill, R.T., 2002, "The Speed of Light and the Einstein legacy: 1905-2005", eprint arXiv:physics/0501051.
Cahill, R.T., 2003a, "QUANTUM FOAM, GRAVITY AND GRAVITATIONAL WAVES", arXiv:physics/0312082 (2003).
Cahill, R.T., 2003b, "GRAVITY AS QUANTUM FOAM IN-FLOW", arXiv:physics/0307003v1.
Cahill, R.T., 2004, "ABSOLUTE MOTION AND GRAVITATIONAL EFFECTS", Apeiron 11, pp53-111 (2004).
Cahill, R.T., 2005, "The Michelson and Morley 1887 Experiment and the Discovery of Absolute Motion", Progress in Physics, 3, pp25-29 (2005).

Cahill, R.T. and K. Kitto, 2002, "Re-analysis of Michelson-Morley Experiments reveals Agreement with COBE Cosmic Background Radiation Preferred Frame so Impacting on Interpretation of General Relativity", arXiv:physics/0205070 (2002).

Consoli, M., 2003, "Relativistic analysis of Michelson-Morley experiments and Miller's cosmic solution for the Earth's motion", eprint arXiv:physics/0310053 (2003).
Consoli, M., 2005, "Precision test for the new Michelson-Morley experiments with rotating cryogenic cavities", eprint arXiv:physics/0506005 (2005).

Consoli, M., and E. Costanzo, 2003a, "The Motion of the Solar System and the Michelson-Morley Experiment", eprint arXiv:astro-ph/0301576 (2003).

Consoli, M., and E. Costanzo, 2003b, Phys. Lett. A318 (2003) 292-299.

Consoli, M., and E. Costanzo, 2004, Nuovo Cim. B119 (2004) 393-410.

Deen, G.W., 2003, "Re-Analyzing Dayton C. Miller's Raw Interferometer Data", Natural Philosophy Alliance Conference, Storrs, Connecticut, (2003).

DeMeo, J., 2001, "Dayton Miller's Ether-Drift Experiments: A Fresh Look", Infinite Energy Magazine, 38, p72-81 (2001). Also: http://www.orgonelab.org/miller.htm

Handschy, M.A., 1982, "Re-examination of the 1887 Michelson-Morley Experiment", Am. J. Phys. 50(11), p987-990 (1982).

Illingworth, K.K., 1927, "A Repetition of the Michelson-Morley Experiment Using Kennedy's Refinement", Phys. Rev. 30, p692-696 (1927).

Michelson, A.A., and E. W. Morley, 1887, "On the Relative Motion of the Earth and the Luminiferous Ether", Am. J. Sci., XXXIV, p333-345 (1887).

Miller, Dayton C., 1928, Astrophys. J, LXVIII(5), p341-402 (1928).

Miller, Dayton C., 1933, "The Ether-Drift Experiment and the Determination of the Absolute Motion of the Earth", Rev. Mod. Phys, 5, p203-242 (1933).

Munera, H.A. 1997, "An Absolute Space Interpretation (with Non-Zero Photon Mass) of the Non-Null Results of Michelson-Morley and Similar Experiments: An Extension of Vigier's Proposal", APEIRON Vol.4 Nr. 2-3, pp77-79 (1997).

Munera, H.A., 1998, "Michelson-Morley Experiments Revisited: Systematic Errors, Consistency among Different Experiments, and Compatibility with Absolute Space", Apeiron 5(1-2), p37-53, (1998).

Rabiner, L.R., 1975, *Theory and Application of Digital Signal Processing*, Prentice-Hall, 1975.

Roberts, Thomas J., 2006, "What is the Experimental Basis of Special Relativity", Usenet FAQ, http://math.ucr.edu/home/baez/physics/Relativity/SR/experiments.html

Sato, 2006, "Interpretation of the slight periodic displacement in the Michelson-Morley experiments", arXiv:physics/0605067 (2006).

Selleri, F., 2000, "On the Anisotropy Observed by Miller and Kennedy & Thorndike" in M. Duffy & M. Wegener, eds., *Recent Advances in Relativity Theory 2: Material Interpretations*, pp. 281-283, Hadronic Press, Palm Harbor (2000).

Shankland, R.S. et al, 1955, "New Analysis of the Interferometer Observations of Dayton C. Miller", Rev. Mod. Phys., 27 no. 2, p167-178 (1955).

Vigier, J.P., 1997, "Relativistic Interpretation (with non-zero photon mass) of the small ether drift velocity detected by Michelson, Morley, and Miller, Apeiron, 4(2), p133 (1997).

Will, Clifford M., 2005, "Special Relativity: A Centenary Perspective", eprint gr-qc/0504085.

# FIGURES WITH CAPTIONS

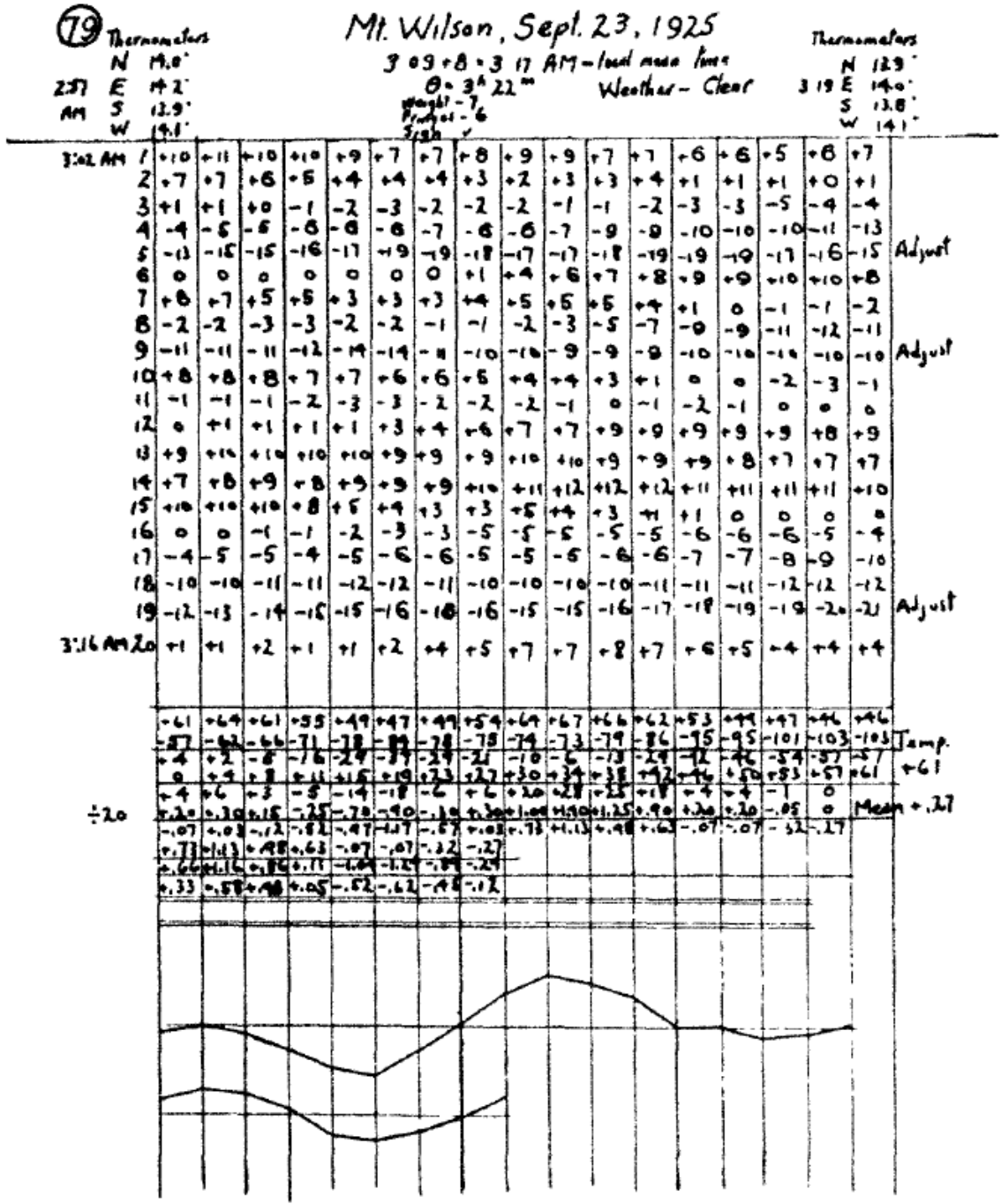

FIG. 8. Form of record of ether-drift observations.

**FIG 1. Figure 8 from Miller (1933), showing a typical page from his logbook.**

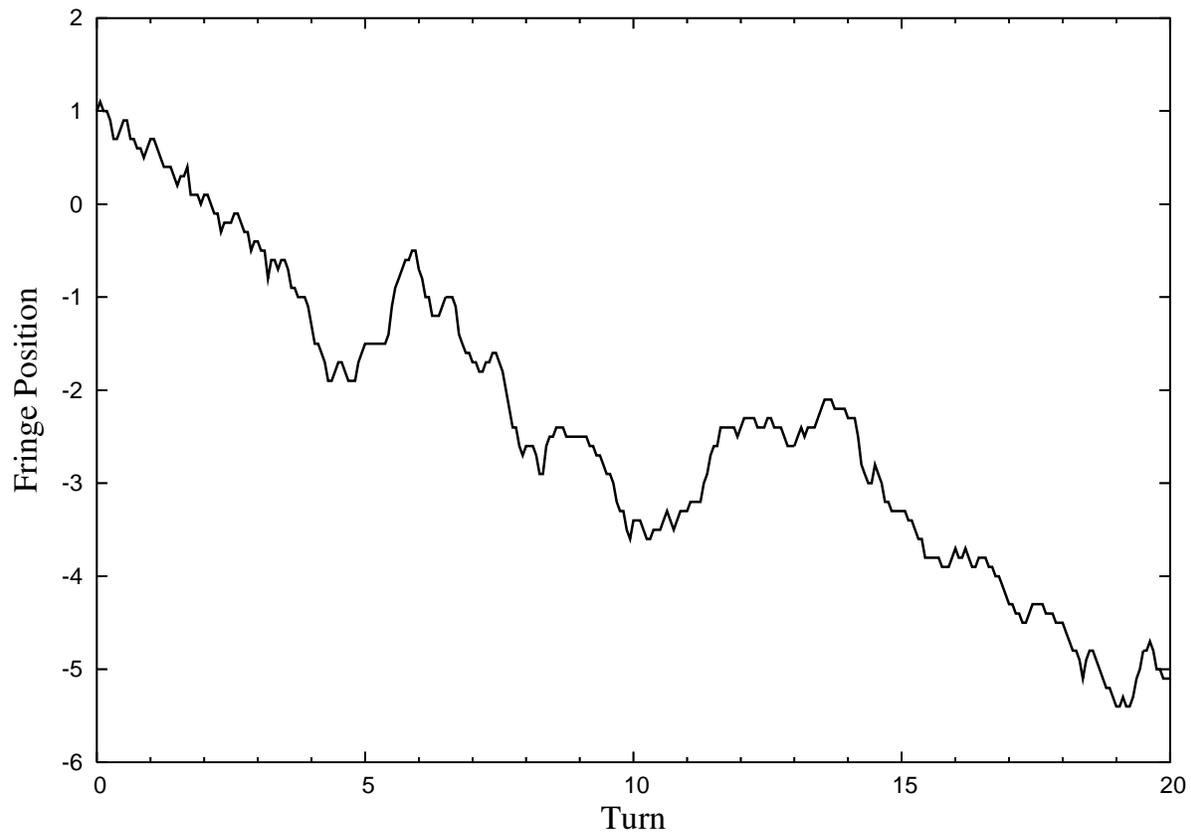

**FIG 2. The raw data from Fig. 1 (adjustments restored).**

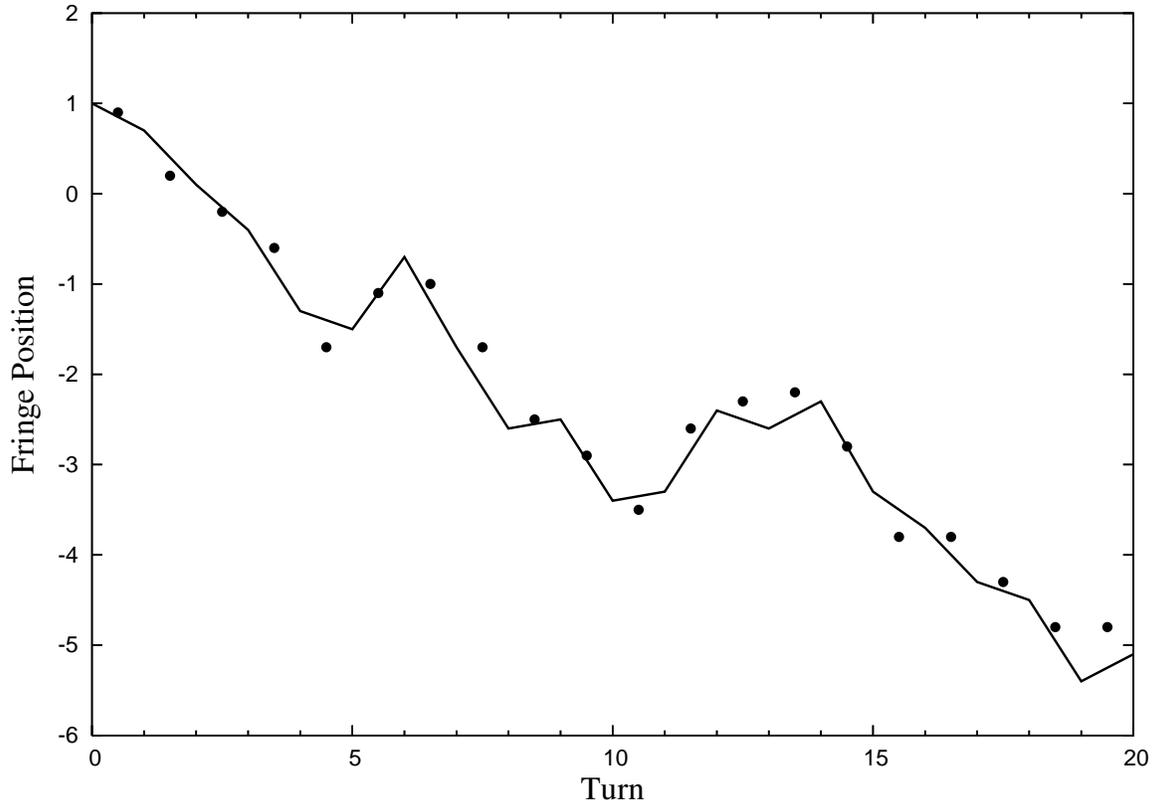

**FIG 3. The assumed-linear systematic drift from the data of Fig. 1.
The lines are between successive Marker 1 values and the points are Marker 9.
These markers are 180 degrees apart, so any real signal has the same value
for every corner and every point – the variations are purely an instrumentation effect.**

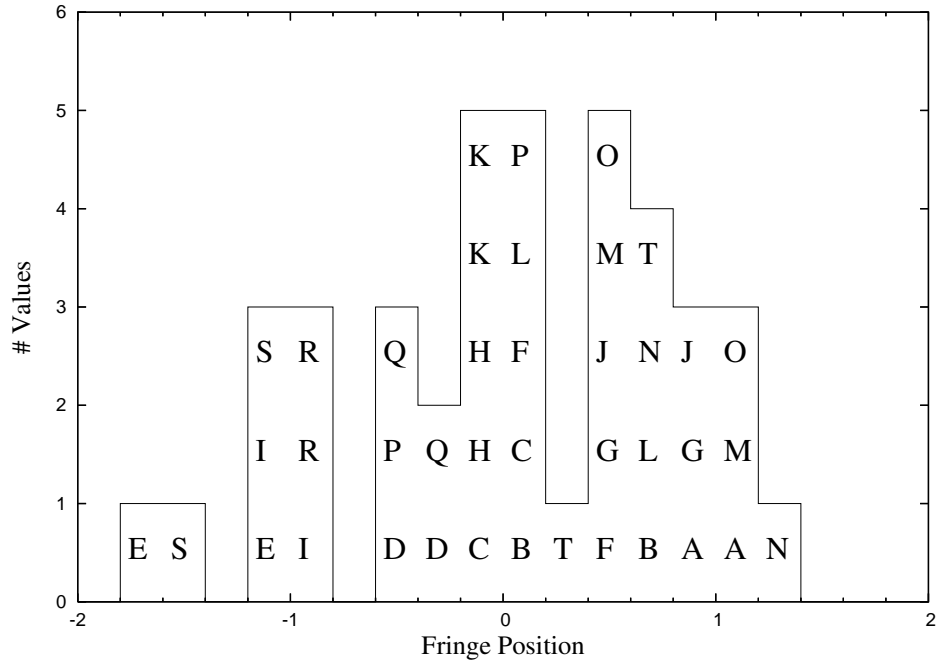

**FIG 4. Histogram of the 40 data points from Markers 1 and 9, augmented with symbols A-T for turns 1-20; the patterns show this histogram is dominated by the systematic drift.**

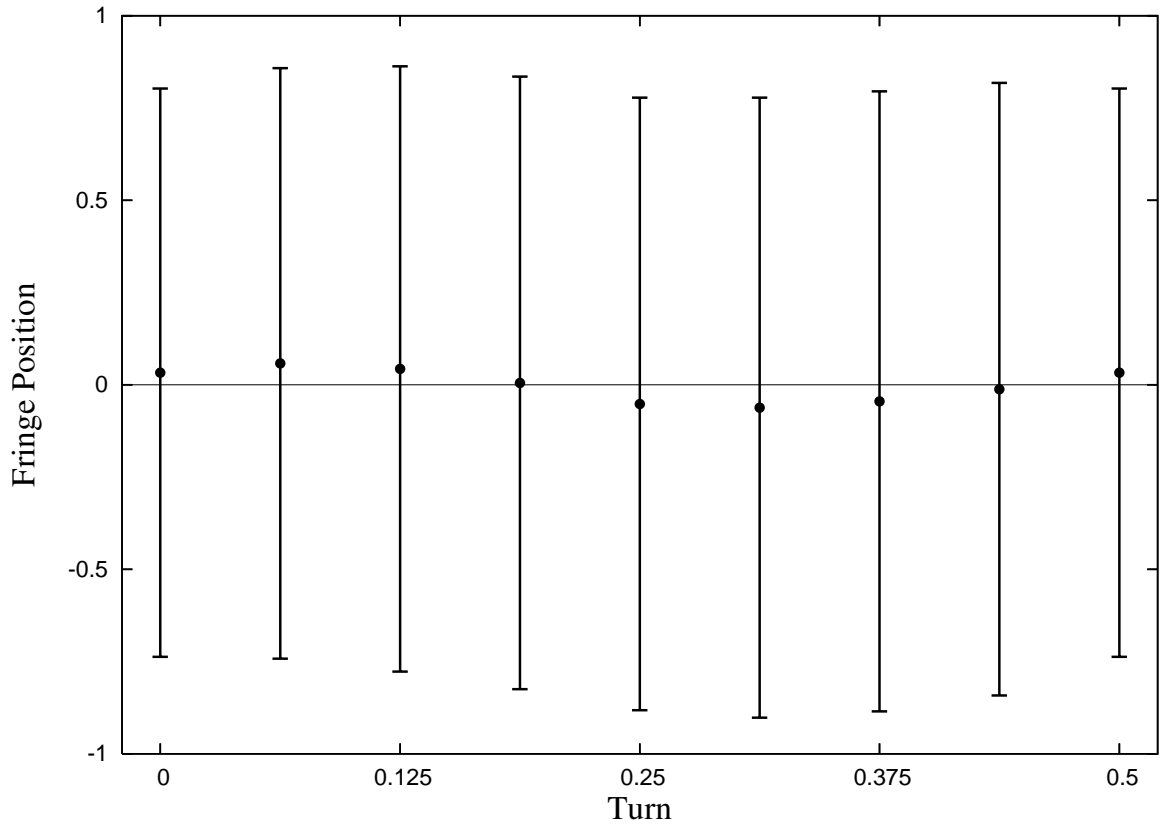

**FIG 5. The plot at the bottom of Fig. 1, with errorbars (see text).**

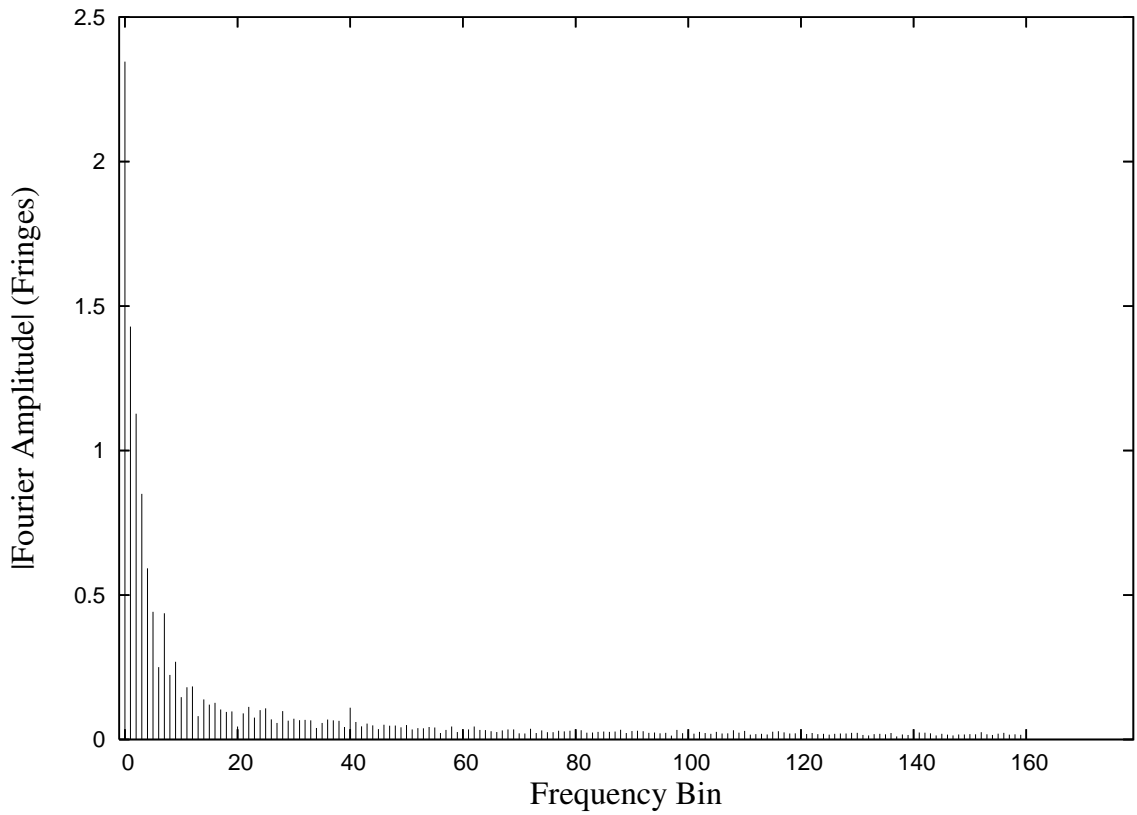

**FIG 6. 320-point DFT spectrum of the data from Fig. 1.
Frequency bin 40 has a period of ½ turn, where any real signal would be.**

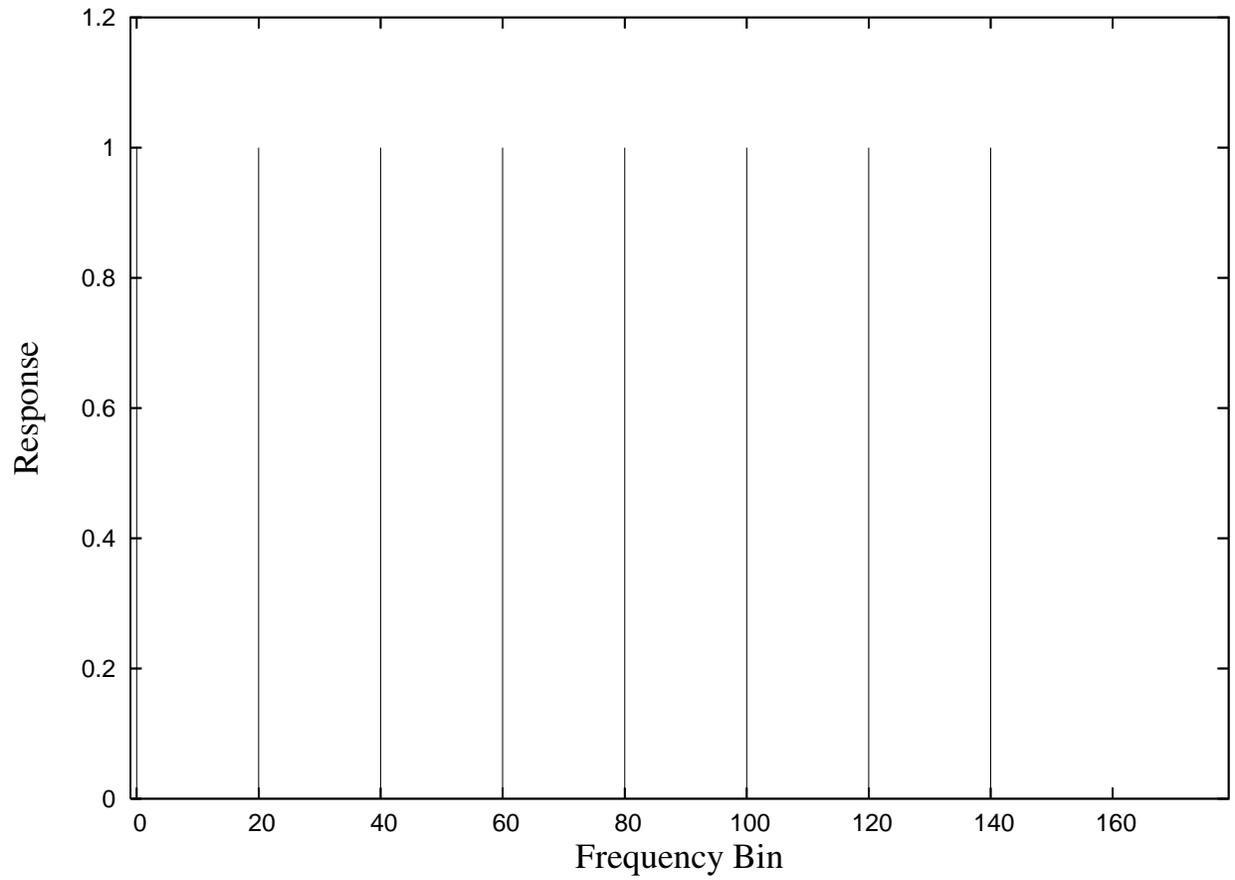

**FIG 7. Frequency Response of the comb filter corresponding to averaging 20 turns.**

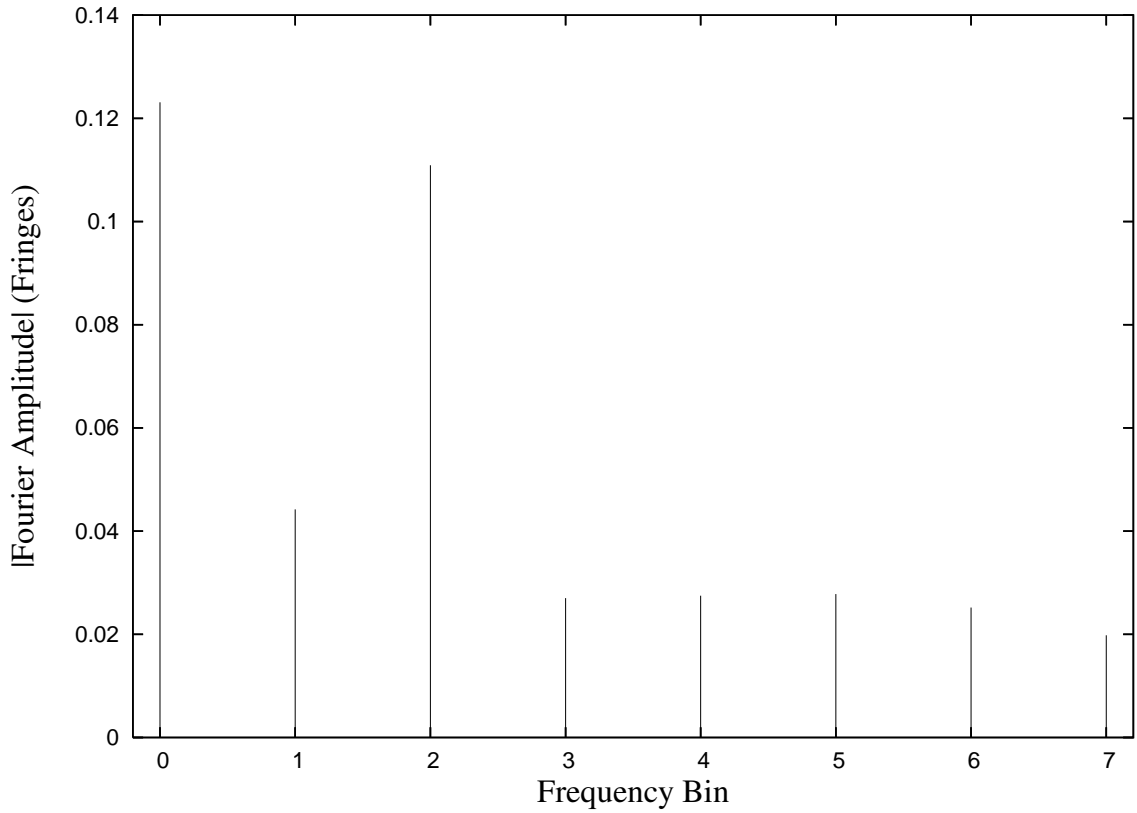

**FIG 8. 16-point DFT spectrum of the data from Fig. 1 after averaging the 20 turns. Frequency bin 2 has a period of ½ turn, where any real signal would be.**

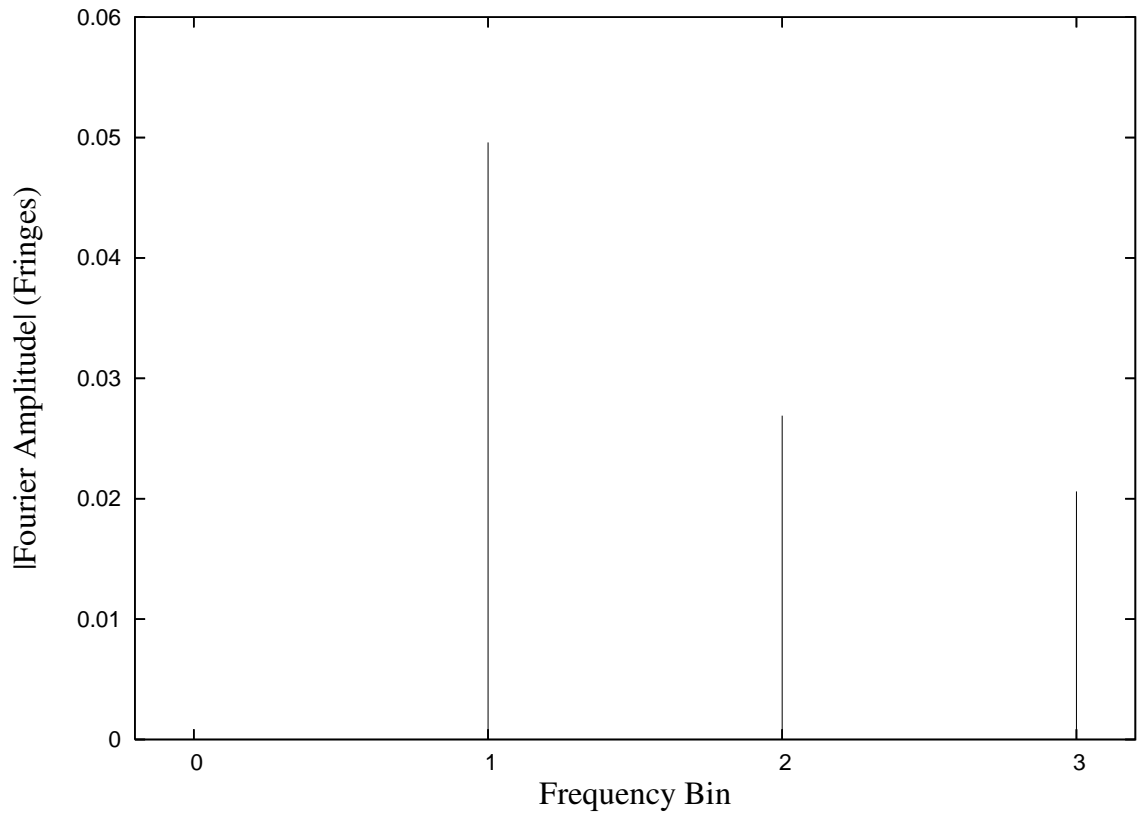

**FIG 9. 8-point DFT spectrum of the data minus the assumed-linear systematic; this corresponds to the final plot of Fig. 1. Any real signal is in frequency bin 1.**

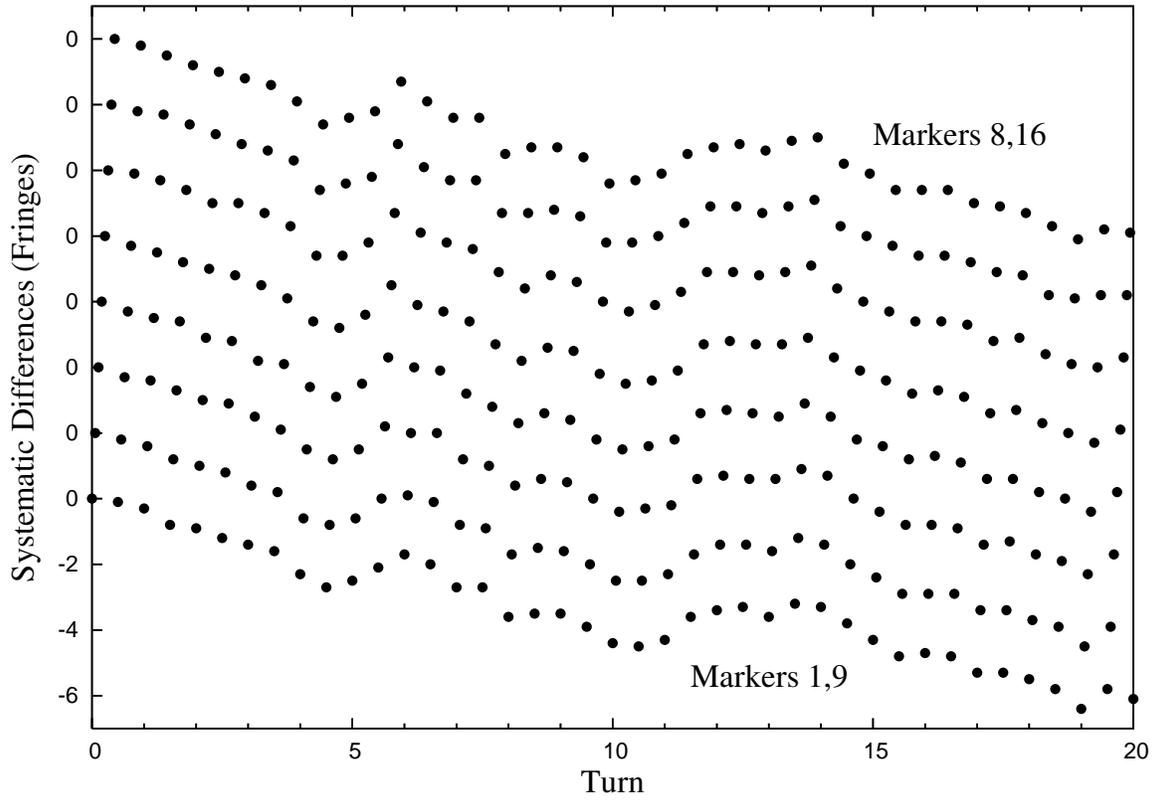

**FIG 10. The measurements of the systematic drift differences for the eight orientations. Because the first turn has been subtracted, any real signal has been completely removed (see text). Origins are offset vertically for clarity.**

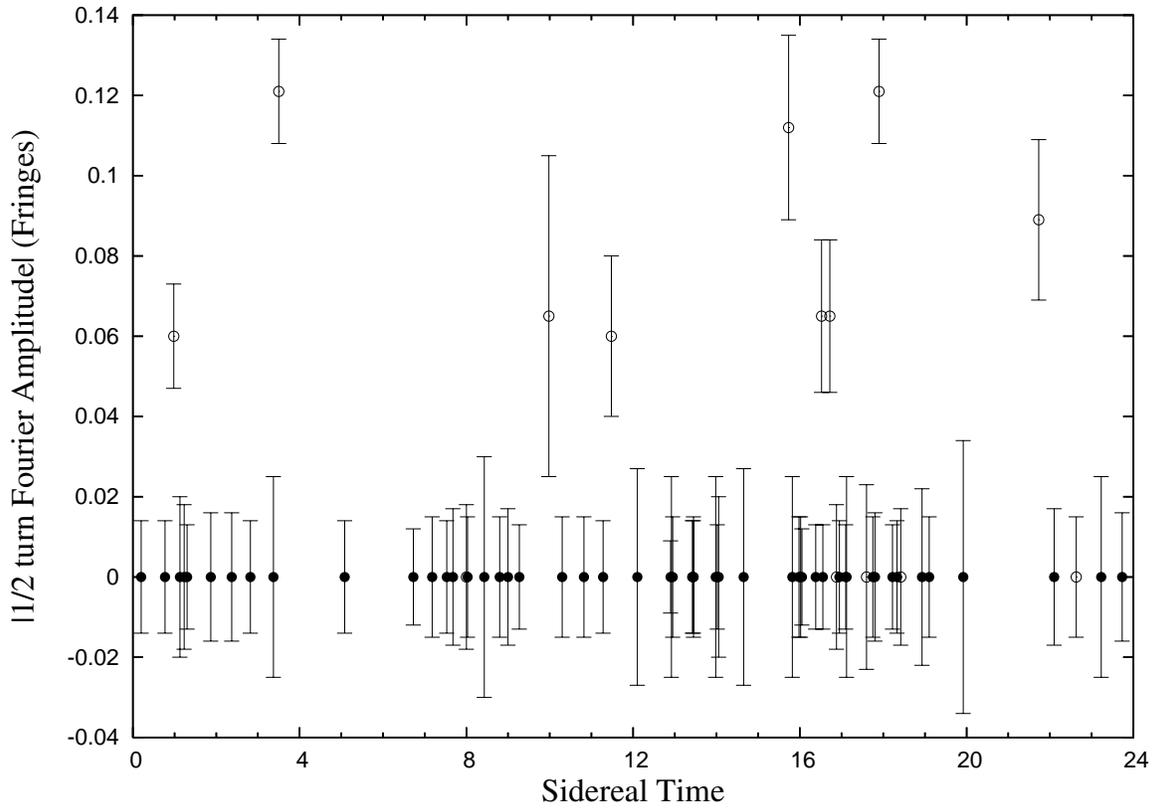

**FIG 11. The ½-turn DFT amplitude, data minus systematic, for 67 Runs. Runs with open circles have five or fewer stable turns (out of 20 turns). The lack of variance around zero is due to the quantization of the data.**

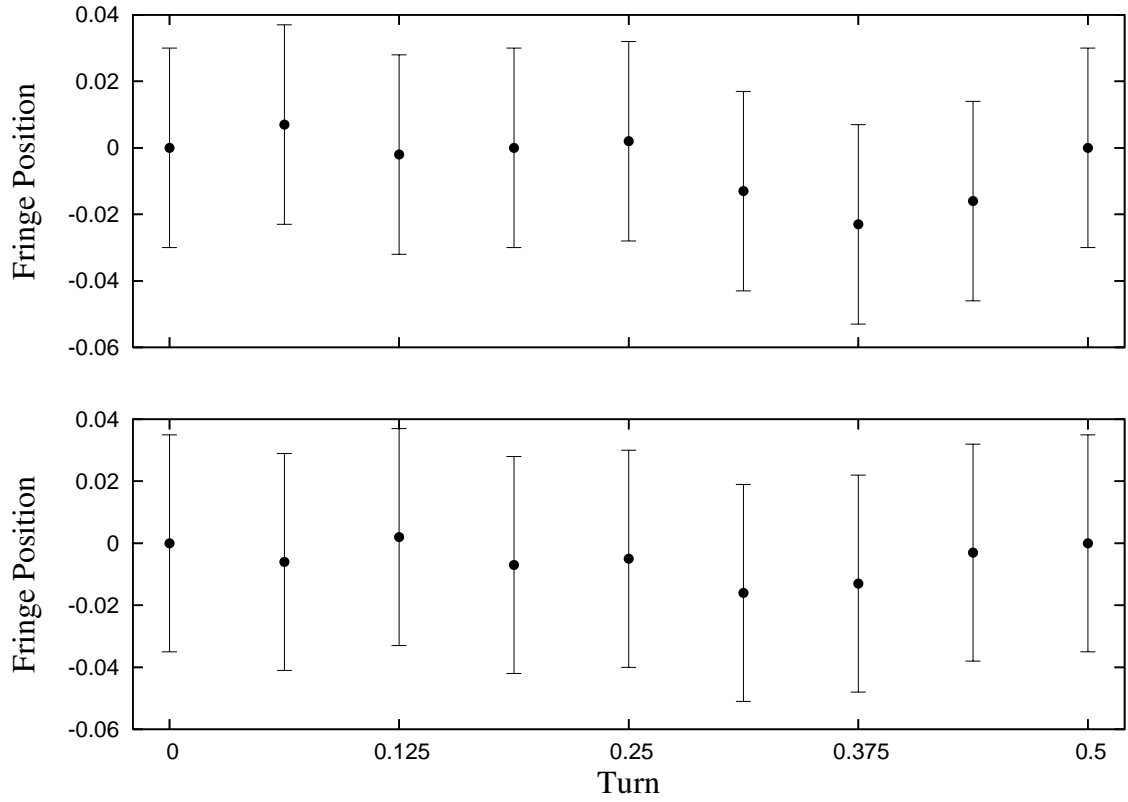

**FIG 12. The Michelson-Morley data, Noon (upper) and P.M. (lower), with errorbars (see text). These errorbars are under estimates.**

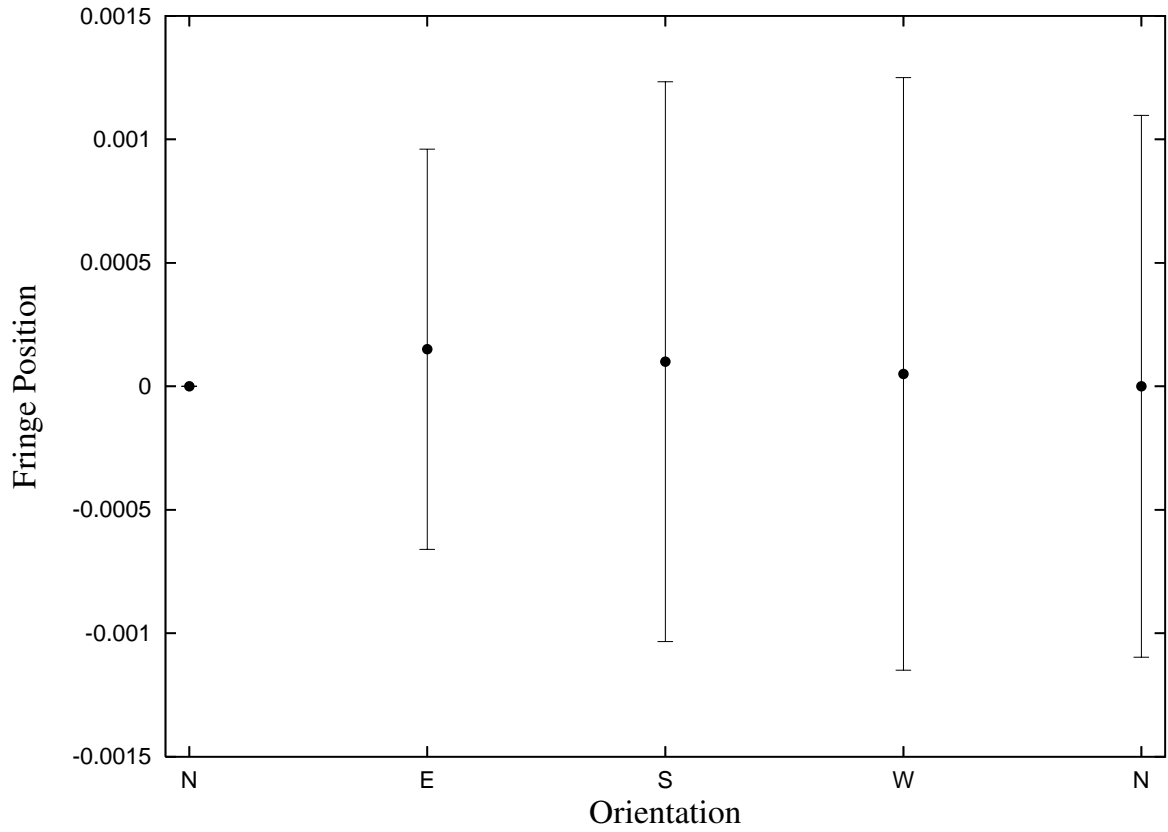

**FIG 13. The Illingworth data minus the assumed-linear systematic model.
See text for an explanation of the errorbars – they are probably under estimates.**